Title: Noncommutative Topological Quantum Field Theory-Noncommutative Floer Homology
Authors: I. P. Zois
Comments: 20 pages, Tex
Report-no: SMAUK-10-05
Subj-class: theoretical physics, mathematical physics, geometric topology, differential geometry, quantum algebra
Journal-ref: 

We present some ideas for a possible Noncommutative Floer Homology. The geometric motivation
comes from an attempt to build a theory which applies to practically every 3-manifold (closed, 
oriented and connected) and not only to homology 3-spheres. There is also a physical motivation: one 
would like to construct a noncommutative topological quantum field theory. The two motivations are closely
related since in the commutative case at least, Floer Homology Groups are part of a certain (3+1)-dim 
Topological Quantum Field Theory.

\documentstyle[12pt,amsfonts,amssymb,amscd]{article}
\begin{document}
\title{Noncommutative Topological Quantum Field Theory-Noncommutative Floer Homology}
\author{Ioannis P. \ ZOIS\thanks{zoisip@cf.ac.uk}\\
\\ 
School of Mathematics, Cardiff University
\\
PO Box 926, Cardiff CF24 4YH, United Kingdom
\\}
\date{}
\maketitle
\begin{abstract}

We present some ideas for a possible Noncommutative Floer Homology. The geometric motivation
comes from an attempt to build a theory which applies to practically every 3-manifold (closed, 
oriented and connected) and not only to homology 3-spheres. There is also a physical motivation: one 
would like to construct a noncommutative topological quantum field theory. The two motivations are closely
related since in the commutative case at least, Floer Homology Groups are part of a certain (3+1)-dim Topological Quantum 
Field Theory.

Classification: theoretical physics, mathematical physics, geometric topology, differential geometry, quantum algebra\\


\end{abstract}

\section{Introduction and Motivation}

This article describes some ideas which emerged during out visit at the IHES in 2002. Since then progress has been slow
so we decided to put these notes on the web hoping that they may attract some attention and someone might sed some light
 on certain interesting we believe issues raised hereby. It is more like a research project than a complete article.\\

Our motivation is twofold: it comes both from geometry and from physics. Let us start with the first, geometry. 
We would like to deepen our understanding on 3-manifolds. Floer Homology is a very useful device since it is the only known 
homology theory which is only \emph{homeomorphism} and \textsl{not homotopy} invariant. Yet computations are particularly 
hard and the theory itself is very complicated; moreover the notorius reducible connections make things even worse 
and at the end Floer Homology Groups are defined only for homology 3-spheres. We would like to have a hopefully simpler 
theory which would apply to a larger class of 3-manifolds. We shall elaborate more on this in the next sections.\\

Our second motivation comes from physics, to be more specific we are interested in \textsl{quantum gravity} and 
\textsl{unifying theories}. 
To begin with, most physicists take the point of view that quantum gravity-which is currently an elusive
theory-\emph{should exist;} the argument in favour of its existence goes as follows (the original argument was
due to P.A.M. Dirac): let us consider Einstein's classical field equations 
which describe gravity (we assume no cosmological constant and we set the speed of light $c=1$):
$$G_{\mu\nu}=8\pi GT_{\mu\nu}$$
In the above equation, $G$ denotes Newton's constant, $T_{\mu\nu}$ denotes the energy-momentum tensor and 
$G_{\mu\nu}$ denotes the Einstein 
tensor which is equal, by definition, to $G_{\mu\nu}:=R_{\mu\nu}-\frac{1}{2}Rg_{\mu\nu}$, where $g_{\mu\nu}$ is the 
Riemannian metric, $R_{\mu\nu}$ 
is the Ricci curvature tensor and $R$ is the scalar curvature. One can see clearly that the RHS of the above
equation,namely the energy-momentum tensor, contains 
mass and energy coming from the other two interactions in nature; mass for instance, consists of
fermions (quarks) and leptons and we know that these interactions (strong and electroweak) are quantized and hence
the RHS of the equation contains \emph{quantized quantities}. So for \emph{consistency} of the equations, the 
LHS, which encodes geometry, \textsl{should also be quantized}.\\

[Here we  would like to mention a small coment as an Aside: one may argue that the LHS may remain 
classical while the RHS may involve the \textsl{average value} of an operator; however such a theory will not be 
essentially different from classical general relativity and probably not qualified to be called quantum gravity, what we 
have in mind is Ehrenfert Theorem from Quantum Mechanics. We think of the above 
field equations as describing, in the quantum level, an actual equality between operators].\\
 
Now the famous and very well-known \emph{Holography Principle}, which has attracted
a lot of attention since 1992 when it was proposed originally by G. 't Hooft (see \cite{thooft}), states that 
quantum gravity should
be a \emph{topological quantum field theory} as defined by Atiyah in \cite{atiyah}. One could expect that to happen, 
even without holography because of the 
following simple fact: given (for simplicity) a closed Riemannian 4-manifold and the Einstein-Hilbert action which contains 
the square root of the scalar curvature of the metric as Lagrangian density, in order to compute the partition function 
of the theory
one would have to integrate over all metrics. It is clear that if one was able to perform this functional integral, 
the result should be a topological invariant of the underlying manifold simply because "there is nothing else left" apart 
from the topology of the Riemannian manifold. We take for brevity the 4-manifold
 to be closed, so Atiyah's axioms for a topological quantum field theory will reduce to obtaining numerical invariants 
and not elements of a vector space associated to the boundary (eg Floer Homology Groups of the boundary 3-manifold).
But here there is an impotant question: the partition
function of the Einstein-Hilbert action on a Riemannian manifold should be a topological invariant, but should it
be a \emph{diffeomorphism} \textsl{or a} \emph{homeomorhism} (or even homotopy) invariant? We know from the stunning work 
of S.K. Donaldson in the '80s (see \cite{donald}) that the DIFF and the TOP categories \textsl{in dimension 4} 
are two entirely different 
worlds (existence of "exotic" ${\bf R}^{4}$'s)! So particularly for the case of 4-manifolds (which is our intuitive idea for 
spacetime) this question is crucial. [We would like to make a remark here: in physics literature the term "topological"
really  means "metric independent" without further specification but for 4-dim geometry, this point is 
particularly important].
We do not have a definite answer on this but it is an issue which in most cases it is not addressed to in the literature;
 however we feel that for quantum physics TOP should be more appropriate as the working category since for example in quantum 
mechanics (solutions of Schrodinger equation) one requires only continuity and not smoothness of solutions at points 
connecting different regions.\\

But this is not enough; if we want a unifying theory of all interactions, we must have other fields present apart from the 
metric (eg gauge fields for electroweak and strong interactions and/or matter fields). We know from the case of the 
\textsl{Quantum Hall Effect} and Bellissard's
work (and others') (see eg \cite{connes}) that the existence of \emph{external fields} "make things \emph{noncommutative}". 
For the
particular case of the QHE the presence of a uniform magnetic field turns the
 Brilluin zone of a periodic crystal from a 2-torus to a noncommutative 2-torus. Further evidence for this phenomenon,
namely the appearence of noncommutative spaces when external fields are present, comes from string theory: the 
Connes-Douglas-Schwarcz article (\cite{cds}) indicates that when a constant 3-form $C$ (acting as a potential) 
of D=11 supergravity is
turned on, M-theory admits additional compactifications on noncommutative tori. Also in string theory, the Seiberg-Witten
article (see \cite{sw}) also discusses noncommutative effects on open strings arising from a nonzero $B$-field. So we 
believe there is good motivation
to try to see what a possible \emph{noncommutative topological quantum field theory} should look like since from what
we mentioned above, it is reasonable to expect that a unifying quantum theory should have some noncommutativity arising from
the extra gauge or other fields present; it should also be a topological quantum field theory since it should contain 
quantum gravity.\\

\section{Topological Invariants for 3-manifolds}

Let us start by recalling some well-known facts from 3-manifold topology: we fix a nice Lie group $G$, 
say $G=SU(2)$; if $M$ is a 3-manifold with fundamental group $\pi _{1}(M)$, 
then the set 
$$R(M):=Hom(\pi _{1}(M),G)/ad(G)$$ 
consisting of equivalence classes
 of representations of the fundamental group $\pi _{1}(M)$ of $M$ onto the Lie 
group $G$ modulo conjugation \textsl{tends to be discrete}.
 If $M$ is a homology  3-sphere, ie $H_{1}(M;{\bf Z})=0$, (this is a safficient
 condition but not in any way necessary), then $R(M)$ has a 
\emph{finite} number of elements and the \textsl{trivial representation} is 
\textsl{isolated}.\\

There is a well-known 1:1 correspondence between the elements of the set $R(M)$
 and elements of the set\\ 

$A(M):=\{$\emph{flat} $G$-connections on $M\}$/(gauge equivalence)\\

The bijection is nothing other than the \emph{holonomy} of the flat 
connections.\\

 Although $R(M)$ \emph{depends on the homotopy type of} $M$, we can get 
\emph{topological invariants} of $M$, ie invariants under 
\textsl{homoeomorphisms}, if we use the moduli space $A(M)$: depending on how we
 ``decorate'' the elements of $A(M)$, namely by giving different ``labels'' to 
the elements of $A(M)$, we can get the following \emph{topological invariants} 
for the 3-manifold $M$:\\

{\bf 1.} \emph{The (semi-classical limit of the) Jones-Witten invariant}.\\
Pick $G=O(n)$ and for each (gauge equivalence class of) flat 
$O(n)$-connection $a$ say on $M$, we have a flat $O(n)$-bundle $E$ over $M$ 
with flat $O(n)$-connection $a$ along with its exterior covariant derivative 
denoted $d_{a}$; now since $a$ is flat, $d_{a}^{2}=0$ and hence we can form the
  \textsl{twisted de Rham complex} of $M$ by the flat connection $a$ denoted
$(\Omega ^{*}(M,E), d_{a})$, where $\Omega ^{*}(M,E)$ denotes smooth 
$E$-valued 
differential forms on $M$. We equip $M$ with a Riemanian metric and we also 
define the \textsl{twisted Laplace operator} by the flat connection $a$ to be:
$\Delta _{a}:=d_{a}^{*}d_{a}+d_{a}d_{a}^{*}$.
Then the
\emph{Ray-Singer analytic 
torsion} $T(M,a)$ is a \emph{non-negative real number} defined by the formula (see \cite{rs}):

$$log[T(M,a)]:=\frac{1}{2}\sum_{i=0}^{3}(-1)^{i}i\zeta '_{\Delta _{i,a}}(0)$$

where $\Delta _{i,a}$ denotes the twisted Laplace operator acting on $i$-forms
and

$$\zeta '_{\Delta _{i,a}}(0):=-\frac{d}{ds}\zeta _{\Delta _{i,a}}|_{s=0}=
log D(\Delta _{i,a})$$

and where we call $D(\Delta _{i,a})$ the $\zeta$-function regularised 
determinant of the Laplace operator $\Delta _{i,a}$ (this is a generalisation 
of the logarithm of the determinant of a self-adjoint operator).\\
The 
$\zeta $-function of the Laplace operator $\zeta _{\Delta _{i}}$
is by definition (for $s\in {\bf C}$):

$$\zeta _{\Delta _{i}}(s):=\sum _{\{\lambda _{n}\geq 0\}}\lambda _{n}^{-s}=\frac{1}{\Gamma (s)}\int_{0}^{\infty}t^{s-1}Tr(e^{-t\Delta _{i}})dt$$

for $Re(s)$ large. Then $\zeta _{\Delta _{i}}$ extends to a meromorphic 
function of $s$ which is analytic at $s=0$.\\

\textsl{The Ray-Singer analytic torsion is independent of the Riemannian metric
 if the twisted de Rham cohomology groups are trivial.}\\

If $M$ is a homology 3-sphere (or any other 3-manifold such that the set $A(M)$
 has finite cardinality), then if we sum-up the Ray-Singer analytic 
torsions of all the flat connections 
(since these are finite in number we know the sum will converge), what we shall get 
as a result is a topological invariant of the 3-manifold which is closely 
related to the ``low energy limit'' (or the semi-classical limit) of the 
\emph{Jones-Witten} (or Reshetikin-Turaev) quantum invariants for 3-manifolds (see \cite{witten}). 
More precisely
 the low energy limit of the Jones-Witten quantum invariants for homology 
3-spheres is a finite 
sum of combinations of the Ray-Singer torsions with the corresponding 
Chern-Simons numbers (ie the integral of the Chern-Simons 3-form over the 
compact 3-manifold $M$) of the flat connections.\\

{\bf 2.} \emph{The Casson invariant}.\\
Let $M$ be a homology 3-sphere and pick $G=SU(2)$. If we choose a 
Hegaard splitting on $M$, then assuming that $R(M)$ is regular (ie that the 1st
 twisted de Rham cohomology groups vanish for all flat connections), then each 
element of $R(M)$ aquires an orientation,
namely a ``label'' +1 or -1. Let us denote by $c_{-}$ (resp $c_{+}$) the number
 of elements of $R(M)$ with orientation -1 (resp +1). Both $c_{-}$ and $c_{+}$ 
depend on the Hegaard splitting chosen but their 
\emph{difference} $c:=c_{-}-c_{+}$ \emph{does not} (in fact it behaves like an index) and this integer $c$ is the 
\emph{Casson invariant} of the 3-manifold $M$. Clearly $c$ is well defined 
since the cardinality of $R(M)$ is \emph{finite} and hence both $c_{-}$ and
$c_{+}$ are finite.\\

{\bf 3.} \emph{Floer Homology Groups}.\\
Again $M$ is a homology 3-sphere (and hence both $R(M)$ and $A(M)$ 
have a finite number of elements); we pick $G=SU(2)$,  we denote by $B(M)$ 
the space of \textsl{all} $SU(2)$-connections on $M$ modulo gauge 
transformations and we denote by $B^{*}(M)$ the \textsl{irreducible} ones (a 
connection is irreducible if its stabiliser equals the centre of $SU(2)$ where 
the stabiliser is the centraliser of the holonomy group of a connection). We 
want to do \textsl{Morse Theory} on the $\infty$-dim Banach manifold $B(M)$:\\ 
(i). We find a suitable ``Morse function'' $I:B^{*}(M)\rightarrow {\bf R}$: 
this is
 the integral over $M$ of the Chern-Simons 3-form
$$I(A)=\frac{1}{8\pi ^{2}}\int_{M}Tr(A\wedge dA+\frac{2}{3}A\wedge A\wedge A)$$
 with a \emph{finite} number of \textsl{critical points}; these are precisely 
the  elements of $A(M)$. This is true since the solutions of the Euler-Lagrange
 equations for the Chern-Simons action are the flat connection 1-forms.\\
(ii). Then each element of $A(M)$ aquires a 
``label'' which is the \emph{Morse index} of the critical point; in ordinary 
finite dim Morse theory  this is equal to the number of negative eigenvalues of
 the Hessian. But the Hessian of the Chern-Simons function is unbounded below 
and we get $\infty$ as Morse index for every critical point. So naive 
immitation of ordinary finite dim Morse theory techniques do not work.\\

Floer in \cite{floer} observed the following crucial fact: pick a Riemannian metric on $M$ and 
considering the noncompact 4-manifold ${\bf R}\times M$ along with its 
corresponding Riemannian metric, a continuous 1-parameter family of connections
 $A_{t}$ on $M$ corresponds to a unique connection ${\bf A}$ on 
${\bf R}\times M$; then, choosing the axial gauge (0th component of the 
connection vanishes), the \textsl{gradient flow} equation 
for the Chern-Simons function $I$ on $M$ corresponds to the \textsl{instanton 
equation} on the noncompact 4-manifold ${\bf R}\times M$:
$$\partial _{t}A_{t}=*F_{A_{t}}\Leftrightarrow F_{{\bf A}}^{+}=0.$$
 Then consider the \textsl{linearised} instanton equation $d_{{\bf A}}a=0$, 
where $a$ is a small perturbation. This operator is not elliptic; we perturb it
to $D_{{\bf A}}=-d_{{\bf A}}^{*}\oplus d_{{\bf A}}^{+}$ to make it elliptic. 
Then the \emph{finite} integer Morse index for each critical point comes as 
the relative (with respect to the trivial flat connection) Fredholm index of 
the perturbed elliptic operator $D_{{\bf A}}$. In this way the moduli space 
$A(M)$ aquires a ${\bf Z}/8$
grading and then we follow ideas from Morse theory:\\ 
(iii). We define the Floer-Morse 
complex using as generators the critical points and the ``differential'' is 
essentially defined by the flow lines of the critical points. Taking the 
cohomology in the usual way we get the \emph{Floer homology groups} of $M$. 
The Euler characteristic of the 
Floer-Morse complex equals twice the Casson invariant (see \cite{donald}).\\

{\bf Remarks:}

(a). The structure Lie group $SU(2)$ can be replaced by another group, say 
$U(2)$.\\
(b). We assumed that all critical points were not only \emph{non-degenerate} 
(i.e. $H^{1}_{A}(M)=0$, this denotes the first twisted de Rham cohomology group
 of $M$ by the flat connection $A$), but in fact \emph{acyclic} (i.e. 
$H^{0}_{A}(M)=H^{1}_{A}(M)=0$). If this is not the case, then the theory just 
becomes more complicated and one has to use \emph{weighted spaces}.\\
(c). One needs a restriction of the form $b^{+}>1$ in order to be able to prove
 independence on the choice of the Riemannian metric (the Riemannian metric 
defines a Hodge star operator whose square equals $1$, hence its eigenvalues 
are $\pm1$; this gives a splittting of the space of 2-forms into positive and 
negative eigenspaces and $b^{+}$ simply denotes the \textsl{positive} part of 
the 2nd Betti number).\\
(d). Reducible connections create more severe problems; this is the main reason
 why people usually work with homology 3-spheres:  apart from having a finite 
number of gauge equivalence classes of flat connections, they have a unique 
reducible connection which is the trivial flat connection which is moreover 
isolated. If one wants to take the reducible connections into account as well, 
then one has to use \emph{equivariant} Floer homology. This is a lot more 
complicated and less satisfactory as a theory since equivariant Floer Homology 
groups may be \textsl{infinite dimensional} and hence there is no Euler 
characteristic for the equivariant Morse-Floer complex; also there is no 
Casson invariant known in this case.\\

 All the above depend crucially on the fact that $R(M)$ (or equivalently 
$A(M)$) has finite cardinality; the most convenient case that this is 
guaranteed is if $M$ is a 
homology 3-sphere. So the question is: what happens if $M$ is such that $R(M)$ 
does not have finite cardinality? Is there a chance to define the 
analogue of the Casson invariant say in this case or even more than that, a 
Floer homology?\\

We believe \textsl{``yes''} and this is precisely the point we are trying to 
develope here.\\

The key idea is the following: we want to replace $R(M)$
by another more stable and better behaving moduli spcace. To do that we use 
as our basis a recent result by David Gabai (see \cite{gabai}): For practically
 any 3-manifold 
$M$ (closed, oriented and connected), the moduli space $N(M)$ of taut codim-1 
foliations modulo coarse isotopy has \emph{finite} cardinality.\\

More concretely: a codim-1 foliation $F$ (through the 
Frobenius theorem this is given by an integrable subbundle $F$ of the tangent 
bundle $TM$ of our 3-manifold $M$) is called \textsl{topologically taut} if 
there exists
a circle $S^{1}$ which intersects transversely all leaves. A codim-1 foliation 
is called \textsl{geometrically taut} if there exists a Riemannian metric on 
$M$ for which all leaves are minimal surfaces (ie they have mean curvature 
zero). One can prove that a codim-1 foliation is geometrically taut if and only
 if it is topologically taut. Foliations in general are very flexible 
structures and the taut foliations are the most rigid ones. Let us call the 
quotient bundle $Q:=TM/F$ the \textsl{transverse bundle} to our foliation.\\

Let $M$ be a Riemannian 3-manifold. Two codim-1 foliations on $M$ are called
\emph{coarse isotopic} if up to isotopy of each one of them their oriented 
tangent planes differ pointwise by angles less than $\pi $. Then Gabai proves 
the following (Theorem 6.15 in \cite{gabai}): Given any closed, orientable, 
atoroidal
3-manifold $M$ with a triangulation, there exists a finite non-negative integer
$n(M)$ such that any taut codim-1 foliation on $M$ is coarse isotopic to one
of the $n(M)$ taut codim-1 foliations. The condition that $M$ should be 
atoroidal may be relaxed as Gabai points out. It is clear that $n(M)$ is the 
cardinality of the Gabai moduli space $N(M)$.\\

The crucial fact is that although the definition of coarse isotopy depends on 
the Riemannian metric, the number $n(M)$ \emph{does not.}\\

Let us emphasise here that although the Gabai moduli space is finite 
practically for any 3-manifold, it may turn out to be \emph{empty} [for example, $S^{3}$ has no taut codim-1 
foliations].\\

The first idea is to try to see if one can immitate the definition of the 
Casson invariant using the Gabai moduli space, namely if we choose a Hegaard 
splitting, can we define a Casson type of invariant?\\

 As a second step we may try to define a Floer type of theory using the Gabai 
moduli space. To do that we need to look for ``labels'' for the elements of the
 Gabai moduli space $N(M)$, ie \emph{invariants for foliations.} There is 
one well-known invariant for codim-1 foliations, the the Godbillon-Vey 
invariant (see for example \cite{candel}) which is the integral over our 
compact 3-manifold $M$ of the
Godbillon-Vey class which for codim-1 foliations on $M$ is a 3-dim real de Rham
 cohomology class defined as follows: let $F$ be a transversely oriented 
codim-1 integrable subbundle of the tangent bundle $TM$ of our closed, oriented and
connected 3-manifold $M$. Locally $F$ is defined by a nonsingular 1-form say 
$\omega$ where $F$ consists precisely of the vector fields which vanish on 
$\omega$ (ie the fibre $F_{x}$ where $x\in M$ equals Ker $\omega _{x}$). The 
integrability condition of $F$ means that $\omega\wedge d\omega =0$. This is 
equivalent to $d\omega=\theta\wedge \omega$ for another 1-form $\theta$. Then 
the Godbillon-Vey class is the 3-dim real de Rham cohomology class 
$[\theta\wedge d\theta ] \in H^{3}(M;{\bf R})$. The problem however with the GV 
invariant is that it is only invariant 
under \textsl{foliation cobordisms} (see \cite{candel}) which is a 
\emph{more narrow} equivalence relation than 
coarse isotopy, hence we may lose the finiteness of the Gabai moduli space 
(equivalently if we use the GV-invariant, we should restrict ourselves to only 
those 3-manifolds with a finite number of taut codim-1 foliations modulo 
foliation cobordisms; unfortunately we do not know if any such 3-manifolds 
exist at all).\\ 
  
Then the idea is to use \emph{noncommutative geometry} techniques in order to 
give labels to the elements of $N(M)$. We intruduced a new invariant for 
foliated manifolds (see \cite{zois}) using indeed 
noncommutative geometry tools, in particular Connes' pairing between cyclic 
cohomology and K-Theory. The foliation has to be transversely oriented with a holonomy invariant transverse measure, 
these restrictions are quite mild. 
Connes' approach to foliations as described in 
\cite{connes} is to complete the 
holonomy groupoid of a foliation to a $C^{*}$-algebra and then study its 
corresponding K-Theory and cyclic cohomology. The invariant in \cite{zois} is 
constructed by defining a canonical K-class in the K-Theory of the foliation 
$C^{*}$-algebra and then pair it with the \textsl{transverse fundamental cyclic
 cocycle} of the foliation. To give a flavour of what that means we describe it
 in the commutative case, ie when the foliation is a fibration, in particular a principal $G$-bundle 
(where $G$ is a nice Lie group): if we have a 
fibration seen as a foliation over a compact manifold (the foliated manifold is
 the \textsl{total space} of the fibre bundle), 
then this transverse fundamental cyclic cocycle is the fundamental homology 
class of the base manifold which is transverse to the leaves=fibres; the 
$C^{*}$-algebra is Morita equivalent to the commutative algebra of functions on
 the base manifold. By the Serre-Swan theorem the K-Theory of this commutative 
algebra coincides with the Atiyah topological K-Theory of the base manifold and
 Connes' pairing reduces to evaluating say Chern classes over the fundamental 
homology class of the base manifold (here we use the 
Chern-Weil theory to go from K-Theory to the de Rham cohomology). The key 
property of the canonical K-class constructed in \cite{zois} is that it takes 
into account the natural action of the holonomy groupoid onto the transverse 
bundle of the foliation. [Note: In some sense this class is similar to the canonical class in $G$-equivariant K-Theory,
 for $G$ 
some Lie group acting freely on a manifold, the situation is more complicated in the foliation case since instead of a 
Lie group we have the holonomy groupoid of the foliation acting naturally on the transverse bundle].
We also need the result that the $G$-equivariant 
K-Theory of the total space of the principal $G$-bundle is isomorphic to the 
topological K-Theory of the quotient by the group action (since this is a 
$G$-bundle, the quotient by the $G$-action is the base manifold). But this 
invariant has not yet 
been properly understood: obviously if it is to be used to define invariants 
for 3-manifolds using the Gabai moduli space it should be invariant under 
coarse isotopy or under a broader equivalence relation.  For the 
moment this point is unclear.\\

Another thing which seems interesting, following what we know from the 
commutative case, is to try to define a \emph{Ray-Singer torsion for foliated 
manifolds} and then try to see if this is invariant under coarse isotopy. In order to define the Ray-Singer 
analytic tosrion one needs a flat connection. For the case of foliations, a flat connection always exists, it is 
our friend the 1-form $\theta$ appearing in the definition of the GV-class; this can indeed be seen in a natural way as
a connection on the transverse bundle (for arbitrary codimension $q$, $\theta$  can be seen as a flat connection on the 
$q$th exterior power of the transverse bundle, this is always a line bundle). This 1-form is sometimes refered to as the
(partial) \emph{flat} Bott connection; it is flat (=closed since this is real valued 
ie Abelian),
 only when restricted to the leaf directions (which justifies the term partial; this is harmless, it can be extended
to a full connection by, for example, using a Riemannian metric).\\

 Yet there seem to
 be two further possibilities here: one can also 
define the \textsl{tangential Laplace operator} and define its corresponding Ray-Singer analytic torsion
(see \cite{moore}, we shall define tangential cohomology 
in the next section). Yet one can use the 
\textsl{Cuntz-Quillen Laplacian} on 
cyclic cohomology (of the foliation $C^{*}$-algebra) defined in \cite{quillen}. In order to define the Ray-Singer analytic 
torsion of the Cuntz-Quillen 
Laplacian one needs a convergence condition because the cyclic complex is unbounded [an idea would be to use what is called
\emph{entire cyclic cohomology} which encorporates an additional analytic convergence condition on the cyclic (co)cycles].
 This Cuntz-Quillen Laplacian enabled Cuntz-Quillen to prove a \textsl{``harmonic 
decomposition''} theorem for cyclic cohomology, in a purely algebraic context, which is analogous to the 
well-known Hodge theorem for the de Rham complex on closed manifolds. 
({\bf N.B:} In the Cuntz-Quillen harmonic decomposition theorem the C-Q 
Laplacian does
 not vanish in the ``harmonic part'' as it happens in the Hodge theorem, it is 
only  \emph{nilpotent} there, but with a suitable simple normalisation we can 
have the normalised C-Q Laplacian vanishing in the harmonic part). That's 
another point which deserves further clarification.\\

The Heitsch-Lazarov analytic torsion in \cite{heitsch} is defined for foliated flat bundles and it does not seem to 
be of any use 
here since it is exactly the flat connections moduli space which we want to replace.\\

{\bf Note:} We tend to think of foliations as generalsising \emph{flat} vector 
bundles: one way to manifest the integrability of a flat connection $a$ say is
to say that its exterior covariant derivative $d_{a}$ has square zero 
$d_{a}^{2}=0$, ie it is a differential. Something similar happens for 
foliations if one considers the ``tangential'' (or ``leafwise'') exterior 
derivative on the foliated manifold which is taking derivatives along the leaf 
directions only; this gives rise to the ``tangential Laplace operator'' 
mentioned above along with the so called \emph{tangential cohomology} 
and it has corresponding \emph{tangential Chern classes} (see \cite{moore}). 
In the above sense tangential cohomology can be seen somehow as a 
generalisation of the twisted de Rham cohomology by a flat connection. Under 
the light of this note the analytic torsion defined by Heitsch-Lazarov in 
\cite{heitsch} has some unsatisfactory properties for our purpose since it is 
a torsion for a foliated flat bundle (namely a flat bundle whose base sapce is
 in addition, foliated, and so the total space carries essentially 3 structures
: the fibration, the foliation where the leaves are covering spaces of the base
 space--flatness--and another foliation which under the bundle projection 
projects leaves to leaves. \\  

The 3rd point which is the most ambitious is to try to define a sort
of Floer homology using the Gabai moduli space. In order to do that
one needs to develop a Morse theory for foliated manifolds. One has at
 first to find a Morse function whose critical points will
be the taut codim-1 foliations. Immitating perhaps naively the Floer
homology case we have two natural candidates for a Morse function:
tangential Chern-Simons forms and Chern-Simons forms for cyclic
cohomology as developped by Quillen not very long ago in \cite{quillen} (that's
a noncommutative geometry tool). The hope is that by using the Gabai
moduli space one might have a chance to avoid the problems with
reducible connections (ie the ``bubbling phenomenon'', see
\cite{donald}) when trying to define Floer homology groups for
3-manifolds which are not homology 3-spheres. There are some more versions of 
Floer homology available but they need some extra structure: a $spin^{c}$ 
structure for the Seiberg-Witten version (and use of the monopole equation 
instead of the instanton equation), or a symplectic structure (as in the 
original Floer attempt) or a complex 
structure (as in the Oszvath-Szabo approach where one uses complex holomorphic 
curves instead of instantons).\\

One of the problems one faces in the above is that there seem to be at least 
three cohomology theories which can describe foliated manifolds (more precisely
 the space of leaves of a foliated manifold): the tangential 
cohomology, the cyclic cohomology of the corresponding $C^{*}$-algebra of the 
foliation and the so-called \emph{Haefliger cohomology} (which has been used in
 the construction of the Heitsch-Lazarov analytic torsion). In ordinary Morse 
theory, given a compact smooth manifold, one considers a real valued function 
(called the Morse function) defined on the manifold and under favourable cases 
one can reconstruct the homology of the manifold by using the flows of the 
critical points of the Morse function. In a would-be Morse theory for foliated 
manifolds one would like to reconstruct the homology of the space of leaves 
using a suitable Morse function, but it is currently unclear which homology of 
the 3 above is more suitable! Moreover the critical points should correspond to
 taut foliations in order to use the Gabai moduli space. We think the above 
challenge is fascinating. Some progress towards a Morse theory for foliated 
manifolds has been already made by Connes-Fack. Also N. Nekrasov et al. work on
 \textsl{noncommutative instantons} may give hints on how to preceed building a
 noncommutative Floer Homology using the Gabai moduli space of isotopy classes  
of taut codim-1 foliations.\\

In order to make some progress towards this direction we need at least 2 tools: a \emph{Morse theory} 
(finite dimesnional at a first stage) for \emph{foliated manifolds} and analytic torsion for foliated manifolds.
 We shall say something more concrete about the first in the next section.\\

\section{Morse Theory for Measured Foliations}

Let $(M,F)$ be a smooth foliation on a closed $n$-manifold $M$ (and $F$ is an integrable subbundle of the 
tangent bundle $TM$ of $M$ where $dimF=p$, $codimF=q$ with $p+q=n$), equipped with a holonomy invariant transverse measure
$\Lambda$ (we need that in order to be able to perform the analogue of "integration along the fibres" which we do for vector or principal
$G$-bundles using the Haar measure which is invariant under the group action). We consider the \emph{tangential cohomology} coming from the 
differential graded complex $(d_{F},\Omega ^{*}(M,F))$, where $d_F$ denotes the \emph{tangential} exterior derivative (namely taking derivatives only
along the tangential (leaf) directions) and $\Omega ^{*}(M,F)$ denotes forms on $M$ with values on the bundle $F$. Due to the integrability of $F$, the tangential exterior derivative is also a differential, namely
$d_{F}^{2}=0$, hence we can take the cohomology of the above complex. We pick a Riemannian metric $g$ on $M$ (which, when restricted to every leaf gives
a Riemmanian metric on every leaf), we consider the adjoint operator $d_{F}^{*}$ and we form the \emph{tangential Laplacian} $\Delta _{F}:=d_{F}^{*}d_{F}+d_{F}d_{F}^{*}$.
We know that a \emph{tangential Hodge theorem} holds, hence Ker [$\Delta _{F}^{k}$] (tangential Laplacian acting on tangential $k$-forms) captures the tangential 
cohomology groups (see \cite{connes}). We denote by $\beta _{k}$ the $k$-th tangential Betti number ($0\leq k\leq p$), where clearly 
$\beta _{k}=dim_{\Lambda}[Ker (\Delta _{F}^{k})]$. We must make an important remark here: this is the Murray-von Neumann dimension defined by Connes using 
the invariant transverse measure, it is finite;
the tangential cohomology groups may be infinite dimensional as linear spaces (see \cite{moore}).\\ 

Now let $\phi$ be a smooth real valued function with domain the manifold $M$. A point $x\in M$ is called a \emph{tangential 
singularity} of $\phi$ if $d_{F}\phi (x)=0$.
A tangential singularity is called a \textsl{tangential Morse singularity} if it is nondegenerate, namely the tangential 
Hessian $d_{F}^{2}\phi (x)$ is nonsingular. The \emph{index} of a t-Morse singularity (in the sequel, "t" stands for
 tangential), is defined
 as the number of "-" signs in the signature of the Hessian (quadratic form) $d_{F}^{2}\phi (x)$. We denote by 
$S_{F}(\phi)$, $S_{1,F}(\phi)$ and $S_{1,F}^{i}(\phi)$
the set of all tangential, t-Morse and t-Morse of index $i$ singularities of $\phi$ respectively. We have that 
$S_{F}(\phi)$ is a closed $q$-submanifold of $M$ which 
is \textsl{transverse} to $F$.
 A \emph{tangential Morse function} is one with only tangential Morse singularities, or equivalently it is a function 
which is a Morse 
function when restricted to every leaf. This definition covers our application for 3-manifolds and the Gabai moduli 
space of taut codim-1 foliations. Yet in general
 it is a rather restrictive definition since there are many interesting foliations with no closed transversals. For that,
 we give the definition of an \emph{almost tangential Morse 
function} $\phi$ w.r.t. the holonomy invariant transverse measure $\Lambda$ to be one for which the set \{$x\in M$ s.t. 
$\phi |L_{x}$ has only t-Morse singularities \}
is $\Lambda$-negligible (where $L_{x}$ denotes the unique leaf through the point $x$).\\

It is then not very hard to see that $c_{k}:=\Lambda (S_{1,F}^{k}(\phi))<\infty$. The main results of Connes and Fack
 in \cite{connes} are the following two theorems:\\
{\bf Theorem 1;}
If $q\leq p$ and for any \emph{good tangential almost Morse function}, one has the \textsl{tangential weak Morse 
inequalities}:
 $$\beta _{k}\leq c_{k}$$

{\bf Theorem 2;}
Under the same assumptions as above, one has the \textsl{tangential Morse inequalities}:
$$\beta _{0}\leq c_{0},$$ 
$$\beta _{1}-\beta _{0}\leq c_{1}-c_{0},$$
$$...,$$
$$\beta _{k}-\beta _{k-1}+...+(-1)^{k}\beta _{0}\leq c_{k}-c_{k-1}+...+(-1)^{k}c_{0}$$ 

These results are proved based on some hard analytic results of Igusa and his \emph{Parametrised Morse Theory} 
(see \cite{igusa}) and 
 by using a Witten type (see \cite{witten}) of \emph{perturbed} \textsl{tangential} Laplacian 
$\Delta _{F,\tau}:=d_{F,\tau}^{*}d_{F,\tau}+d_{F,\tau}d_{F,\tau}^{*}$
by a good tangential almost Morse function $\phi$: $d_{F,\tau}:=e^{-\tau\phi}d_{F}e^{\tau\phi}$, where $\tau$ is a 
positive real parameter.\\

We must explain the term good tangential almost Morse function: following the pioneering work of K. Igusa 
(see \cite {igusa}) and his \emph{Parametrised Morse Theory},
Connes and Fack managed to prove the above results not only for tangential Morse functions but for a larger class of
 real valued smooth functions with domain
a foliated manifold: a good tangential almost Morse function $\phi$ is a \emph{generalised} (namely it can contain 
Morse singularities and 
\textsl{birth-death} singularities), almost tangential Morse function which is \emph{generically unfolded}. The last 
requirement means that there exist 
\textsl{normal forms}
describing the function in a neighbourhood of a birth-death singularity. A birth-death singularity is a degenerate 
tangential singularity (i.e. tangential Hessian 
vanishes) for which the restriction of the map $x\mapsto (d_{F}(\phi)(x),det[d_{F}^{2}(\phi)(x)])$ has rank $p$ at $x$.\\

Let us make the following remarks here: Connes and Fack, before proving the tangential Morse inequalities, they prove 
that \textsl{every} measured foliation with 
$q\leq p$ has 
at least one good tangential almost Morse function; their proof is based on an astounding theorem due to K. Igusa: it 
was a well-known fact that a generic smooth
real valued function on a closed manifold has only nondegenerate critical points; however a generic 1-parameter family
 of real valued smooth functions has in addition
birth-death points where critical points are created or canceled in pairs. A multi-parameter family has a zoo of 
complicated singularities; K. Igusa proved that more 
complicated singularities can be avoided: for any foliation on a closed manifold it is always possible to find a
 smooth real valued function such that singularities
 associated with the critical points of its restriction to every leaf are at most of degree 3! Clearly we think of a 
foliation as a more complicated parametrised family
of manifolds than a fibre bundle: the family of manifolds (leaves-they correspond to the tangential directions) is 
parametrised by the space of leaves 
(corresponds to the transverse directions); in a fibre bundle we have a family of manifolds (fibre) parametrised by the
 base manifold.

It is not true that any measured foliation with $q\leq p$ has a tangential Morse function, namely the foliations with
 tangential Morse functions are rather
special (they must have a closed transversal); taut foliations nevertheless, which is what we are mostly interested in,
 do have, by definition, a complete
closed transversal).\\

If we denote by $A(M,F)$, $J(M,F)$ and $R(M,F)$ the sets of tangential almost Morse functions, tangential generalised 
Morse functions and 
tangential generalised Morse functions which are generically unfolded respectively, then the good tangential almost
 Morse functions are those in the intersection of
$A(M,F)$ and $R(M,F)$ (clearly the 3rd set is a subset of the second). The hard piece due to K. Igusa is to prove that
 for a closed $M$ and an $F$ with
$codimF\leq dimF$, the set $J(M,F)$ is nonempty.

For $\phi$ a good tangential almost Morse function, we have that the critical manifold $S_{F}(\phi)$ is a $q$-dim 
submanifold of $M$ transverse to $F$, the set of 
tangential Morse singularities of index $i$ $S_{1,F}^{i}(\phi)$ is also a $q$-submanifold of $M$ transverse to $F$ and 
open inside the critical manifold (but not closed
in $M$ in general) and the set of tangential birth-death singularities $S_{2,F}^{i}(\phi)$ of index $i$ of $\phi$ is a 
closed $(q-1)$-submanifold of the critical manifold
and it is both in the closure of $S_{1,F}^{i}(\phi)$ and of $S_{1,F}^{i+1}(\phi)$.\\

\emph{Acknowledgements:} I would like to thank Tierry Fack for drawing my attention to tangential Morse Theory and J.L.
 Heitsch for giving me a copy of his joint article with C. Lazarov.  Moreover
I wish to thank Alain Connes and Kim Froysov for useful discussions. I have also been benefited from graduate 
lecture courses on Floer Homology by Simon Donaldson, Peter Kronheimer and others during my Oxford days. Finally I want 
to thank IHES for providing excellent working conditions and a very stimulating atmosphere.\\

\end{document}